# XR: Enabling training mode in the human brain

| | |
|---|---|
| **Sébastien Lozé** | **Philippe Lépinard** |
| Epic Games | Associate professor |
| Montreal, Canada | Université Paris-Est Créteil, France |
| sebastien.loze@epicgames.com | philippe.lepinard@u-pec.fr |

## ABSTRACT

The face of simulation-based training has greatly evolved, with the most recent tools giving the ability to create virtual environments that rival realism. At first glance, it might appear that what the training sector needs is the most realistic simulators possible, but traditional simulators are not necessarily the most efficient or practical training tools. With all that these new technologies have to offer; the challenge is to go back to the core of training needs and identify the right vector of sensory cues that will most effectively enable training mode in the human brain. Bigger and Pricier doesn't necessarily mean better.

Simulation with cross-reality content (XR), which by definition encompasses virtual reality (VR), mixed reality (MR), and augmented reality (AR), is the most practical solution for deploying any kind of simulation-based training. The authors of this paper (a teacher and a technology expert) share their experiences and expose XR-specific best practices to maximize learning transfer.

## ABOUT THE AUTHORS

**Sebastien Loze :** Starting his career in the modeling and simulation community more than 15 years ago, Sébastien has focused on learning about the latest simulation innovations and sharing information on how experts have solved their challenges. He worked on the COTS integration at CAE and the Presagis focusing on Simulation and Visualization products. More recently, Sebastien put together simulation and training teams and strategies for emerging companies like CM Labs and D-BOX. He is now the Simulations Industry Manager at Epic Games, focusing on helping companies develop real-time solutions for simulation-based training.

**Philippe Lepinard**: Former military helicopter pilot and simulation officer, Philippe Lépinard is now an associate professor at the University of Paris-Est Créteil (UPEC). His research is focusing on playful learning and training through simulation. He is one of the founding members of the French simulation association.

# XR: Enabling training mode in the human brain

| Sébastien Lozé | Philippe Lépinard |
| --- | --- |
| Epic Games | Associate professor |
| Montreal, Canada | Université Paris-Est Créteil, France |
| sebastien.loze@epicgames.com | philippe.lepinard@u-pec.fr |

**INTRODUCTION: The evolution of simulation-based training**

**Early adopters**

The area that saw the first adoption and development simulation-based training was the training of aircraft pilots. The simulation cost was so high that only pilots and their equipment represented a risk high enough to invest in a tool so sophisticated and expensive. With advances in technology and increasing accessibility, the community opened the potential to train operators and experts in a larger scope of works. Medical personnel, truck operators, first responders, military, robotic operators, and construction equipment operators, among others, are now able to leverage simulators in the context of their training. These two factors, technological advances and diversification of audience, combined are forcing the training and simulation community to reinvent itself and identify disruptive technologies in order to scale up quickly without introducing new costs.

**Bring the trainee to the simulator**

Around the year 2010, many simulation providers went from the model of providing simulators to their customers on site at their place of employment to a model of providing training at centralized training centers. Customers would visit the training centers and rent the necessary hours of training on these simulators. The main driver for such a business model evolution was to lower the total cost of ownership of these traditional simulators while still keeping up with the evolving technology. The center could invest in new technology, and the customer did not have to keep upgrading, providing and training operators for the equipment, etc. For the specific use cases of full flight simulators, where the regulation associations impose specific requirements, the training center model is a good approach.

**Bring the simulator to the trainee**

While the training center model solved some problems, many of the community players need solutions more pragmatic and efficient to train their personal, operators, and drivers. One example would be a company that is situated far from the nearest training center. Another is a niche industry for which custom training is needed. Instead of bringing the trainees to the simulator, we propose a return to the paradigm of bringing the simulators to the trainees. Advances in technology have made the necessary hardware much more affordable, development tools are now much easier to use, and the advent of cloud-based technology gives us a new tool for developing, delivering, and evaluating training. Thanks to the flexibility of cloud-based VR and AR systems, trainees around the world can have access to training. By bringing the training environment on site, curriculum can be adapted, customized, and updated as needed. This paradigm also reduces overall cost by eliminating the costs associated with travel.

**The technology evolution**

In the early days of VR technology, in order to display an image which was close enough to reality in a 3D scene graph, we had to either pre-process it (eliminating a truly interactive experience) or generate images in real time as the viewer turns his/her head (real-time rendering). We know from the film industry that a frame rate of at least 24 fps (frames per second) is required to trick the human brain well enough to accept a moving environment; in XR, 30 fps is considered the minimum acceptable rate. To meet a 30 fps rate with real-time rendering in earlier days of this field, all the visual cues that make a rich, convincing environment--shadows, reflections, complex surfaces--had to be eliminated. Basically, we ended up with a very sterile world, a "synthetic" training environment. New features and technologies are constantly emerging to provide new means, tools, and approaches to simulation-based training with XR. Perhaps most importantly right now, training schools don't have to choose between visual realism, high frame rate, and true-to-life behaviors of simulated objects. A simulation developer can choose any of these goals in developing training for a particular task, or even all three.

## 1. ENABLING TRAINING MODE WITH XR

While we can mimic reality with XR technology, it is not necessarily advisable to do so for a training environment. A too-real environment, besides being time-consuming and costly to produce, can adversely impact learning with overloading. "There is a limit to how much information people can process simultaneously, and this impacts how information is stored. Too much information, or too difficult a task, presented in an ill-considered or unstructured way, can result in cognitive overload for a learner." (Reedy, 2015).

To determine the best level of XR realism for a given training task, we consider XR in the context of the Reality-Virtuality continuum of Milgram and Kishino (1994a, 1994b, Figure 1) and the Reality, Virtuality and Mediality Taxonomy of Mann (2002, Figure 2). These models identify possible combinations of real and virtual elements in order to change the perception of the world by adding, altering or deleting information.

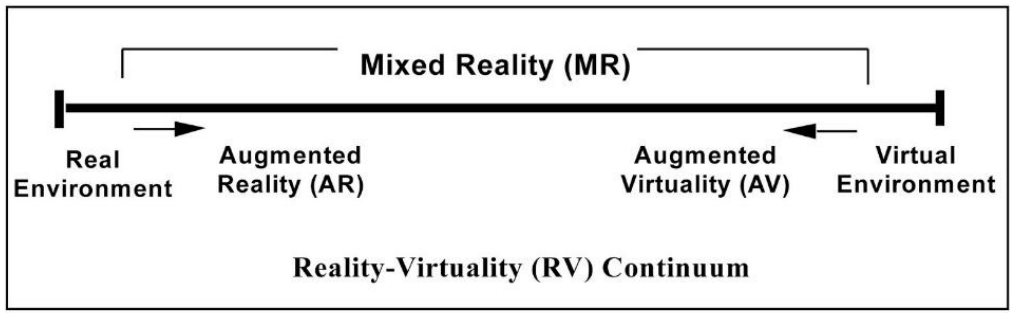

*Figue 1. Reality–virtuality continuum (Milgram and Kishino, 1994)*

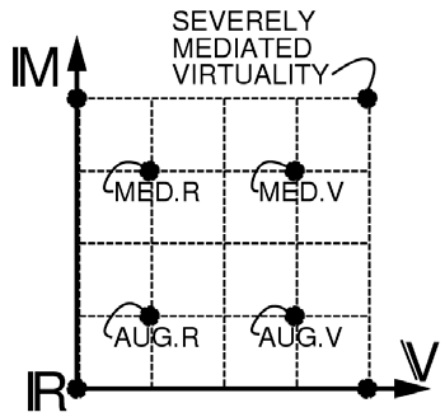

*Figure 2. Reality, Virtuality and Mediality Taxonomy (Mann, 2002)*

These capabilities allow us to consider a parallel pedagogical continuum by offering learners the training device most suited to the targeted learning. This adaptation can then be expressed through an individualization of the training courses. In other words, this adaptation and contextualization of information make it possible to offer learners only scenarios that they will be able to reasonably accomplish at their current skill level.

For example, the technique of a holding pattern in aeronautics can be learned first in a deliberately impoverished virtual world, and then in the real world with an augmented reality device embedded on board and configurable according to the level of the learner and the educational objectives of the trainer.

The strength of the XR is therefore as much (if not more) in the reduction or adaptation of complexity than in the search for the most faithful representation of the real world, always too complex to cognitively apprehend by novice professionals or learners. Cognitive overload can lead to a blocking of learning or even a negative transfer of learning that will be harmful for the following of the training course.

In the context of virtual reality headset, this negative transfer can also come from "simulation sickness" due to a cognitive conflict between vision and the inner ear. These considerations are discussed later in this paper.

Besides these physiological constraints, and in order to properly integrate XR into training courses, didactic and pedagogical engineering must evolve to create links between the new technological capabilities offered by XR and the zone of proximal development concept, that is to say, the margin of development that an individual actually has. "This margin of development is explored by asking an individual to solve problems of a stage higher than the stage in which he finds himself with the help of an adult who plays the role of mediator "(Raynal and Rieunier, 2012, p.510). In our context, we can then imagine an artificial mediator either manually triggered by the learner or the instructor, or in the form of an artificial intelligence coupled with physiological sensors that automatically propose a representation of the world adapted to the mental state and the physics of the learner.

The fidelity of the task doesn't necessarily require a strong sensory and digital immersion during the first stages of its learning, and the human interpersonal relationships in formation (companionship, for example) can limit the uses of certain modalities of XR envisaged by the designers.

These facets of learning with XR have yet to be fully explored. As more institutions develop XR training programs specific to their needs, we shall perhaps see innovative solutions around impoverished-to-rich world progression, serving of learning experiences based on margins of development, and mediator-assisted learning programs.

## 2. XR CHALLENGES AND SOLUTIONS

XR, and particularly VR, has long been embraced by such organizations as the Department of Defense for training military personnel. As VR technology constantly evolves, new challenges appear each year. Here, we discuss the known challenges, and how to overcome them to build an effective simulation-based training system.

### VR content considerations

Creating content for VR is similar in many ways to developing content for a movie or game, but different in many significant ways.

- Scale all objects in the VR world as 1:1 with reality.
- Consider the angles at which viewers will be able to see objects when traversing through the environment.
- Use less costly lighting solutions whenever possible. With a *static* lighting solution, the effects of lights are precomputed, the equivalent of "painting" the lighting and shadows permanently onto objects. With a *dynamic* lighting solution, the lighting is updated at every frame. Static lighting is faster, and less expensive computationally, than dynamic.
- When characters are moving around together in a VR world, precomputed shadows won't work, and characters will appear to float above the ground. An inexpensive solution is a blob or capsule shadow, which follows the character around and casts a roughly blob-shaped shadow on the ground and other surfaces.
- Keep materials and textures as basic as possible. Matte finishes (as opposed to shiny) will look better with static lighting. Transparency and translucency are expensive, so use them sparingly.
- Don't create smoke, fluids, etc. as visual effects unless these are critical to the training exercise. There are ways to fake water and smoke with inexpensive materials, so consider using these instead of their cinematic-quality counterparts.

These are a few of the most important considerations when making a VR world that will play back smoothly at 30 fps or above. Fortunately, most of them argue for greater simplicity in the creation of VR worlds.

**Simulation sickness**

Some users of HMD devices experience a phenomenon known as *simulation sickness*, where simply watching through VR glasses induces headaches, dizziness, and other temporary symptoms. This, of course, ruins the experience for the viewer and makes XR-based training difficult, if not impossible.

Simulation sickness is caused by the fact that visualizing in VR is not the same as viewing reality. Even if the viewer logically knows he/she is safe in an VR world, cognitive conflicts between visual perception and the inner ear cause a physical reaction.

The following are some broad guidelines for making the XR experience safe for all viewers:

- Maintain the HMD's native frame rate, which is 60-120 fps depending on the hardware. This requires optimization of the scene.
- Design the VR environment with dimmer lights and less saturated colors than you would use in 2D media like a traditional video game or movie.
- In developing the simulation, set it up so the viewer has control of his/her view at all times. This means you should not use the following: a cinematic type of camera view, where the simulator and not the trainee chooses what to view; a walking/bobbing head in view, as in a first-person game; a camera shake to indicate an explosion or other disturbance. Similarly, set and maintain the Field of View to match the headset, and don't use cinematic storytelling devices like Depth of Field and Motion Blur. Let the viewer see what they see.
- Movement should accelerate at the same rate at all times. Don't limit initial acceleration then increase it later in the experience.
- Test the simulation on the people who will be getting the training, not the developers. If you don't have trainees handy, test on other people at your company, anyone other than developers. Developers are often too accustomed to VR and XR to be accurate indicators for simulation sickness in the general public.

This is not a complete list of all the considerations for simulation sickness but does provide a starting point.

Despite a significant and permanent improvement of VR technologies, it is surprising that the practices carried out with such helmets (games, professional applications, artistic experiments, etc.) do not specify more formally the known risks of this kind of motion sickness, which can persist for up to 24 hours after the simulation session (Merhi and al., 2007, Paquette and Bélanger, 2015) even for people in good health and having undergone a selection process, as with aircraft pilots.

**CONCLUSION**

XR is not neutral. Its pedagogical integration into an organization (and therefore in a specific culture) must be thought out well in advance, and in a systemic way, in order to avoid a naive technological determinism as was sometimes the case with full-scale simulators, and to facilitate adoption in fields that would benefit greatly from its use.

**BIBLIOGRAPHY AND REFERENCES**

Borg, M., Johansen, S.S., Krog, K.S., Thomsen, D.L., Kraus, M. (2013), Using a Graphics Turing Test to Evaluate the Effect of Frame Rate and Motion Blur on Telepresence of Animated Objects. *GRAPP/IVAPP*.

Elhelw, M., Nicolaou, M., Chung, A., Yang, G.-Z., Atkins, S. (2008), A Gaze-Based Study for Investigating the Perception of Visual Realism in Simulated Scenes, *ACM Transactions on Applied Perception*, Vol.5, No.1.

Mann, S. (2002), Mediated Reality with implementations for everyday life, *Presence Connect*, http://wearcam.org/presence_connect/.

Merhi, O., Faugloire, E., Flanagan, M., Stoffregen, T. (2007), Motion Sickness, Console Video Games, and Head-Mounted Displays, Human Factors: The Journal of Human Factors and Ergonomic Society, Vol.49, No.5, pp.920-934.

Meyer, A., Loscos, C. (2003), Real-Time Reflection on Moving Vehicles in Urban Environments, *VRST '03 Proceedings of the ACM symposium on Virtual reality software and technology*, pp.32-40, Osaka, Japan.

Milgram, P., Kishino, F. (1994a), A taxonomy of mixed reality visual displays, *IEICE Transactions on Information Systems*, Vol.77, No.12, pp.1321-1329.

Milgram, P., Takemura, H., Utsumi, A., Kishino, F. (1994b), Augmented Reality: A class of displays on the reality-virtuality continuum, *SPIE Telemanipulator and Telepresence Technologies*, Vol.2351, pp.282-292.

Paquette, E., Bélanger, D.-C. (2015), *Le mal du simulateur : un frein à l'apprentissage de la prise de décision en conduite d'un véhicule de police ?*, Canadian Journal of Learning and Technology, Vol.41, No.2.

Pausch, R., Crea, T. (1992), A Literature Survey for Virtual Environments: Military Flight Simulator Visual Systems and Simulator Sickness, Computer Science Report, No.TR92-25, http://www.cs.cmu.edu/~stage3/publications/93/journals/PRESENCE/militaryFlight/paper.html.

Perroud, B., Régnier, S., Kemeny, A., Merienne, F. (2017), Model of realism score for immersive VR systems, *Transportation Research Part F: Traffic Psychology and Behaviour*, pp.1-14.

Raynal, F., Rieunier, A. (2012), *Pédagogie, dictionnaire des concepts clés : apprentissage, formation, psychologie cognitive*, Bielsko-Biala : Esf Éditeur.

Reedy, G. (2015), Using Cognitive Load Theory to Inform Simulation Design and Practice, *Clinical Simulation in Nursing*, No.11, pp.355-360.

Vygotski, L. (1985), *Pensée et langage*, Paris : Editions sociales.